\magnification \magstep1
\input amssym.def
\input amssym.tex
\bigskip
\centerline{\bf Bogoliubov Transformations in Black-Hole Evaporation} 
\bigskip
\centerline{A.N.St.J.Farley and P.D.D'Eath}
\bigskip
\smallskip
\centerline{Department of Applied Mathematics and Theoretical Physics,
Centre for Mathematical Sciences,} 
\smallskip
\centerline{University of Cambridge, Wilberforce Road, Cambridge CB3 0WA,
United Kingdom}
\medskip
\centerline{Abstract}
\smallskip
\noindent
In previous papers and letters on quantum amplitudes in black-hole
evaporation, a boundary-value approach was developed for calculating 
(for example) the quantum amplitude to have a prescribed slightly 
non-spherical configuration of a massless scalar field $\phi$ on a
final hypersurface $\Sigma_F$ at a very late time $T{\,}$, given 
initial almost-stationary spherically-symmetric gravitational and
scalar data on a space-like hypersurface $\Sigma_I$ at time 
$t=0{\,}$.  For definiteness, we assumed that the gravitational 
data are also spherically symmetric on $\Sigma_{F}{\,}$.  
Such boundary data can correspond to a classical solution 
for the Einstein/scalar system, describing gravitational collapse 
from an early low-density configuration to a nearly-Schwarzschild 
black hole.  This approach provides the quantum amplitude 
(not just the probability) for a transition from an initial 
to a final state.  For a real Lorentzian time-interval $T{\,}$, 
the classical boundary-value problem refers to a set of 
hyperbolic equations ({\it modulo} gauge), and is badly posed.  
Instead, the boundary-value approach of the previous letters
and papers requires (following Feynman) a rotation into the complex:
$T\rightarrow{\mid}T{\mid}\exp(-i\theta)$, for
$0<\theta\leq\pi/2{\,}$, of the time-separation-at-infinity $T{\,}$.  
The classical boundary-value problem, for a complex solution of the 
coupled nonlinear classical field equations, is expected to be
well-posed for $0<\theta\leq\pi/2{\,}$.  For a locally-supersymmetric 
Lagrangian, containing supergravity coupled to supermatter, the
classical Lorentzian action $S_{\rm class}$, a functional of the
boundary data (which include the complexified $T$), yields a quantum 
amplitude proportional to $\exp(iS_{\rm class})$, apart from possible 
loop corrections which are negligible for boundary data with
frequencies below the Planck
scale.  Finally (still following Feynman), one computes the Lorentzian 
quantum amplitude by taking the limit of $\exp(iS_{\rm class})$ as 
${\,}\theta\rightarrow 0_{+}{\,}$.  In the present paper, a connection 
is made between the above boundary-value approach and the original 
approach to quantum evaporation in gravitational collapse to a black
hole, {\it via} Bogoliubov coefficients.  This connection is developed 
through consideration of the radial equation obeyed by the (adiabatic)
non-spherical classical perturbations.  When one studies 
the probability
distribution for configurations of the final scalar field, based on
our quantum amplitudes above, one finds that this distribution can
also be interpreted in terms of the Wigner quasi-probability
distribution for a harmonic oscillator.\par 
\medskip
\noindent
{\bf 1. Introduction }
\medskip
\indent
In previous letters and papers [1-3], we calculated the quantum
amplitude for a given configuration of the massless scalar field
$\phi$ on a space-like hypersurface $\Sigma_F$ at a very late time
$T$, measured at spatial infinity, given (for simplicity) that both
the gravitational and scalar field configurations $(h_{ij},\,\phi)_I$ 
on an initial hypersurface $\Sigma_I$ are exactly spherically
symmetric, and that the gravitational data $h_{ijF}$ on $\Sigma_F$ are 
also exactly spherically symmetric.  Here, $h_{ij}=g_{ij}$ denotes the 
intrinsic spatial metric $(i,j=1,2,3)$, while ${\,}g_{\mu\nu}$ denotes the
4-dimensional space-time metric $(\mu,\nu=0,1,2,3)$.  Mathematically, 
this calculation involves analytic continuation of the quantum
amplitude into the lower complex $T$-plane.  In [2,3] on the 
`complex approach' and the `spin-0 amplitude' in black-hole evaporation, 
this was phrased in terms of a rotation of $T$ into the complex: 
$T\rightarrow{\mid}T{\mid}\exp(-i\theta)$, where
$0<\theta\leq\pi/2{\,}$.  Provided that the appropriate system of
coupled Einstein/massless-scalar classical field equations is strongly
elliptic [4], up to gauge, for $0<\theta\leq\pi/2{\,}$, the complex 
{\it classical} boundary-value problem, corresponding to the above 
quantum amplitude, should be well posed.  In the extreme case 
$\theta =\pi/2{\,}$, the boundary data have been rotated so as to 
provide a boundary-value problem for a real Riemannian 
(positive-definite) 4-metric $g_{\mu\nu}{\,}$, coupled to a real 
scalar field $\phi{\,}$.  The field equations are 
'elliptic {\it modulo} gauge', for which one expects good behaviour 
of the boundary-value problem, including analyticity of the solutions.\par
\smallskip
\indent 
It is assumed throughout that the original bosonic theory, such as
Einstein gravity with a minimally-coupled massless scalar field, is
embedded in a locally-supersymmetric theory containing supergravity
with supermatter.  In this case, the smallest such theory [5] contains 
$N=1$ supergravity with its spin-3/2 gravitino, a {\it complex} scalar 
field and its massless spin-1/2 partner.  Only for locally-supersymmetric
theories does one expect that the loop terms 
${\,}A_{0},A_{1},A_{2},\ldots{\,}$ in a semi-classical expansion of
the quantum amplitude 
${\rm Amp}\sim(A_{0}+\hbar A_{1}+\hbar^{2}A_{2}+\ldots{\,})
\exp(-I_{B}/\hbar)$ will be finite [6-8].  Here, for our bosonic
boundary data above, $I_{B}$ is the classical 'Euclidean action' of
the boundary-value solution of the coupled Einstein and bosonic-matter 
field equations.  For simplicity, in our example, at present the only 
non-zero boundary data allowed are a real Riemannian 3-metric $h_{ij}$ 
and a real scalar field $\phi{\,}$.  Note that, for dimensional
reasons, the loop corrections will only be significant for boundary
data which contain frequencies at and beyond the Planck scale.\par 
\smallskip
\indent
In [3], the quantum amplitude was then evaluated semi-classically
for $\theta >0{\,}$, and finally the limit of the amplitude was taken as
${\,}\theta\rightarrow 0_{+}{\,}$, to obtain the Lorentzian amplitude 
(that is, the amplitude for $T$ real), following Feynman's
$+i\epsilon$ prescription [2].  In particular (Eq.(4.16) of [3]), we 
computed the imaginary part of the (necessarily) complex spin-0 classical 
action by taking the contribution from the poles along the
real-frequency axis.  The discrete set of real frequencies 
${\,}\sigma_{n}=n\pi/{\mid}T{\mid}{\,}$, for ${\,}n=1,2,3,\ldots ,{\,}$ 
was found to dominate in late-time field configurations.  So far,
however, we have made no explicit mathematical connection with the
usual theory of black-hole evaporation [9-12].  We rectify this
omission in this paper, by considering Bogoliubov transformations 
between (real) massless spin-0-field modes propagating in a 
slowly-varying Schwarzschild-like solution, such as appears in Sec.2
of [3] for adiabatic modes.  Such Vaidya-like approximate classical 
solutions will subsequently be treated in more detail [13].\par
\smallskip
\indent
The original derivation of black-hole radiance involved the
calculation of Bogoliubov transformations between initial and final 
quantum states.  Because of the time-dependence of the gravitational 
collapse, an initial basis of (linearised) positive- and negative-frequency
massless-scalar modes (for example) differs from a final basis.  By
completeness, there exists a (Bogoliubov) transformation between the
initial and final bases.  When this transformation mixes positive and
negative frequencies, as in the black-hole case, one interprets the
result physically in terms of particle creation.\par
\smallskip
\indent
The connection between the approach used in [3] and the approach using
Bogoliubov coefficients arises from the radial equation satisfied
by the (adiabatic) perturbations.  The real (spin-dependent) potential
in the radial equation vanishes sufficiently rapidly at spatial
infinity that the radial functions near infinity (for each spin $s$)
are superpositions of complex exponentials with complex coefficients; 
see Eq.(3.3) of [3] for the case $s=0{\,}$.  In the approach of [3],
regularity conditions for the fields at small radius imply that the 
radial functions are real.  An early-time basis of modes,
corresponding to waves incoming from past null infinity, can be
related, by completeness, to a late-time basis, which is a
superposition of waves outgoing at future null infinity.  It is then 
possible to relate, on a common late-time space-like hypersurface, the 
Bogoliubov coefficients to the complex coefficients above, which arose 
in the boundary-value method, since the Bogoliubov coefficients are 
independent of space-like surface.  The usual (Bogoliubov) approach
gives a density matrix and probabilities for final configurations, but
not the phase information present in a quantum amplitude. \par
\smallskip
\indent
 Working instead with the boundary-value approach [3], we have, as
seen above, a pure state, and can evaluate 
probabilities as ${\,}{\mid}{\rm amplitude}{\mid}^2{\,}$.  Thus, our
approach exhibits a large qualitative difference, as compared to the
usual approach. Correspondingly, there 
should be some differences between the physical predictions based on the 
Bogoliubov approach and on the present approach.\par
\smallskip
\indent
In Sec.2 of the present paper, after a review of the basic properties
of Bogoliubov coefficients, we begin the above process of relating
them, in our problem, to the complex coefficients $z_{n\ell}$ of
Eq.(3.3) of [3].  Sec.3 treats particle emission rates.  In Sec.4, we
consider the resulting probability distribution, in terms of the
steady-state Bogoliubov coefficients.  Sec.5 contains a brief
Conclusion.  The Appendix relates this probability distribution to the
Wigner quasi-probability distribution function for harmonic
oscillators.\par
\medskip
\noindent
{\bf 2. Bogoliubov transformations}
\medskip
\indent
Consider first the (nearly-)Lorentzian collapse problem.  We assume
the existence of a spherically-symmetrical Lorentzian-signature
'reference' or 'background' metric in the form 
$$ds^{2}{\;}{\,} 
={\;}{\,}-{\,}e^{b(t,r)}dt^{2}{\,}+{\,}e^{a(t,r)}dr^{2}{\,} 
+{\,}r^{2}(d\theta^{2} +\sin^{2}\theta{\;}d\phi^{2}){\quad}.\eqno(2.1)$$
In this context, it is conventional to define the 'mass function' 
$m(t,r)$ by 
$$\exp\Bigl(-a\bigl(t,r\bigr)\Bigr){\;}{\,} 
={\;}{\,}1-{{2m(t,r)}\over {r}}{\quad}.\eqno(2.2)$$ 
In the boundary-value problem outlined above, we write
$(\gamma_{\mu\nu}{\,},\Phi)$ for the 'background' spherically-symmetric
metric and scalar field, and $\nabla_{\mu}$ for the background
covariant derivative.\par
\smallskip
\indent
Now consider a classical linearised solution $\phi^{(1)}$ of the 
massless scalar wave equation 
$$\nabla^{\mu}\nabla_{\mu}{\,}\phi^{(1)}{\;}{\,} 
={\;}{\,}0{\quad}.\eqno(2.3)$$
\noindent
Suppose first that one is given two spherically-symmetric Cauchy
surfaces in the space-time [14], an initial Cauchy surface 
${\cal S}^{-}$ and a final Cauchy surface ${\cal S}^{+}$.  Then,
subject to regularity and spatial fall-off conditions, one can expand 
$\phi^{(1)}$ in terms either of a basis 
$\{f_{\omega^{\prime}\ell m}(x)\}$ of mode solutions Eq.(2.12,16) of [3]
adapted to ${\cal S}^{-}$, or of a basis $\{p_{\omega\ell m}(x)\}$ 
adapted to ${\cal S}^{+}{\,}$:
$$\eqalignno{\phi^{(1)}(x){\;}{\,}&
={\;}{\,}\sum^{\infty}_{\ell =0}{\quad}\sum^{\ell}_{m=-\ell}{\;}
\int^{\infty}_{0}{\;}d\omega^{\prime}{\;}{\,}
\Bigl[c_{\omega^{\prime}\ell m}{\;}f_{\omega^{\prime}\ell m}(x){\,}
+{\,}{\rm c.c.}\Bigr]{\quad},&(2.4)\cr
\phi^{(1)}(x){\;}{\,}&
={\;}{\,}\sum^{\infty}_{\ell =0}{\quad}\sum^{\ell}_{m=-\ell}{\;}
\int^{\infty}_{0}{\;}d\omega{\;}{\,}
\Bigl[b_{\omega\ell m}{\;}p_{\omega\ell m}(x){\,} 
+{\,}{\rm c.c.}\Bigr]{\quad}.&(2.5)\cr}$$
\noindent
The c-number complex coefficients $\{c_{\omega^{\prime}\ell m}\}$ are
a set of position-independent 'Fourier amplitudes' labelling the
configuration of the field on ${\cal S}^{-}$, while the 
$\{b_{\omega\ell m}\}$ refer correspondingly to ${\cal S}^{+}$.  
On ${\cal S}^{-}$, the $\{f_{\omega^{\prime}\ell m}\}$ are an 
orthonormal, complete, family of complex solutions of the wave
equation, which contain only positive frequencies
$(\omega^{\prime}>0)$.  Here, in addition to making use of the 
spherical symmetry of the background 4-geometry, one adopts a simple 
definition of 'positive-frequency' in regions where the background 
gravitational and scalar fields are approximately static.  Similarly, 
on ${\cal S}^{+}$, the $\{p_{\omega\ell m}\}$ are an analogous family 
of positive-frequency solutions $(\omega >0)$.\par
\smallskip
\indent
In the presence of a future event horizon, ${\cal H}^{+}$, the final
Cauchy surface ${\cal S}^{+}$ (say) will typically need to cross 
${\cal H}^{+}$. This is the case in the familiar treatments of
black-hole evaporation [9-12], which are based on Bogoliubov
transformations.  Of course, given our analytic-continuation strategy, as
developed in earlier papers [2,3], with the time-interval
at infinity taken to be $T={\mid}T{\mid}\exp(-i\theta)$ for 
$0<\theta\leq\pi/2{\,}$, one has effectively no event horizon in the
nearly-Lorentzian problem.  Rather, one expects the boundary-value
problem to be strongly elliptic, and so to resemble the real
Riemannian boundary-value problem, with no obstruction across any
'horizon'.  Thus, in our case, one can replace the usual Cauchy surface
$S^{+}$, which in the centre of the space-time lies close to the
Lorentzian curvature singularity, with the final surface $\Sigma_{F}$ 
above, on which the final data $(h_{ij}{\,} \phi)_{F}$ are posed.  One 
can take the intrinsic geometry of $\Sigma_{F}$ to be nearly flat, 
with $\Sigma_{F}$ diffeomorphic to ${\Bbb R}^{3}$.  The
time-separation from the initial surface $\Sigma_I$ to $\Sigma_F{\,}$, 
as measured at spatial infinity, is $T={\mid}T{\mid}\exp(-i\theta)$, with
${\mid}T{\mid}$ very large.  A possible choice for a hyper-surface
which crosses $\cal{H}^+$ smoothly is a constant 'time' slice in a
Painlev\'e-Gullstrand-like coordinate system, modified slightly to
account for the slow change of the black-hole mass due to emission;
these coordinates are stationary and non-singular across $\cal{H}^+$
and have been used in a tunnelling interpretation of black-hole
evaporation [15,16 ].\par
\smallskip
\indent
Thus, loosely speaking, we shall consider a 'Bogoliubov transformation
to a smooth surface, long after the Lorentzian singularity'.  In the
same context, one may say that 'the singularity is simply by-passed
in the analytic continuation'.  That is, there should be an analytic
complexified  classical solution to the boundary-value problem, which
only reaches a singular boundary precisely at Lorentzian signature
$(\theta =0)$.  One thus expects to have the possibility of
circumventing Lorentzian singularities by deforming time-intervals 
suitably into the complex, much as one avoids singularities of
functions $f(z)$ in the ordinary complex $z$-plane by deforming contours.\par 
\smallskip
\indent
Suppose that the surfaces $\Sigma_{I}$ and $\Sigma_{F}$ are taken to
be spherically symmetric.  Then, the bases above may be chosen to have the
form
$$\phi_{\omega\ell m}(x){\;}{\,} 
={\;}{\,}N(\omega){\;}{{R_{\omega\ell}(r)}\over{r}}{\;} 
e^{-i\omega t}{\;}Y_{\ell m}(\Omega)\eqno(2.6)$$
\noindent
in terms of the spherical harmonics ${\,}Y_{\ell m}(\Omega){\,}$ [17], 
where $N(\omega)$ is a normalisation factor, and where positive frequency
corresponds to ${\,}\omega >0{\,}$.  On $\Sigma_I{\,}$, one chooses the
large-$r$ behaviour of $R_{\omega\ell}(r)$ such that 
$\phi_{\omega\ell m}(x)$ is proportional to $e^{-i\omega v}$ at large 
$r{\,}$, where ${\,}v=t+r^*{\,}$.  Thus, the 
$\{f_{\omega^{\prime}\ell m}\}$ are purely ingoing at infinity.  At 
$\Sigma_F{\,}$, one requires that $\phi_{\omega\ell m}(x)$ be 
proportional to $e^{-i\omega u}$ at large $r{\,}$, where 
${\,}u=t-r^*{\,}$; thus, the $\{p_{\omega\ell m}\}$ are purely
outgoing at infinity.  We repeat the reasonable assumption for the 
collapse problem, that the initial spatial 3-geometry and scalar field 
(the 'star') can be taken to be approximately static.  Typically, the 
initial Cauchy surface $\Sigma_I$ would be taken to be in the early 
nearly-static region, whereas $\Sigma_F$ would be at a time long after 
the collapse, where the background geometry would be approximately
that of a Vaidya solution with slowly-varying mass [13].\par
\smallskip
\indent
The normalisation factor $N(\omega)$ is determined through use of the
natural inner product [11]
$$(\phi_{\omega\ell m},{\;}
\phi_{\omega^{\prime}\ell^{\prime}m^{\prime}}){\;}{\,}
={\;}{\,}-{\,}i\int_{\Sigma}{\;}d\sigma^{\mu}{\;}
\bigl(\phi_{\omega\ell m}{\,}
{\nabla}_{\mu}\phi^{*}_{\omega^{\prime}\ell^{\prime}m^{\prime}}{\;} 
-{\;}\phi^{*}_{\omega^{\prime}\ell^{\prime}m^{\prime}}{\;}
{\nabla}_{\mu}\phi_{\omega\ell m}\bigr){\quad}.\eqno(2.7)$$
\noindent
For a pair of classical solutions 
$\phi_{\omega\ell m}{\,},{\,}
\phi_{\omega^{\prime}\ell^{\prime}m^{\prime}}{\,}$, this inner product 
is independent of the particular choice of (asymptotically-flat) 
space-like hypersurface $\Sigma{\,}$.  The surface element 
$d\sigma^{\mu}$ in Eq.(2.7) is evaluated in the background
space-time.  The inner product (   ,  ) has the properties 
$$\eqalignno{(\phi_{\omega \ell m}{\,},{\,}\lambda{\,}
\phi_{\omega^{\prime}\ell^{\prime}m^{\prime}}){\;}{\,}&
={\;}{\,}\lambda^{*}{\,}(\phi_{\omega\ell m}{\,},{\,}
\phi_{\omega^{\prime}\ell^{\prime}m^{\prime}}){\quad},&(2.8)\cr
(\lambda{\,}\phi_{\omega\ell m}{\,},{\,}
\phi_{\omega^{\prime}\ell^{\prime}m^{\prime}}){\;}{\,}&
={\;}{\,}\lambda{\,}(\phi_{\omega\ell m}{\,},{\,}
\phi_{\omega^{\prime}\ell^{\prime}m^{\prime}}){\quad},&(2.9)\cr
(\phi_{\omega\ell m}{\,},{\,}
\phi_{\omega^{\prime}\ell^{\prime}m^{\prime}})^{*}{\;}{\,}&
={\;}{\,}(\phi_{\omega^{\prime}\ell^{\prime}m^{\prime}}{\,},{\,} 
\phi_{\omega\ell m}){\quad},&(2.10)\cr}$$ 
\noindent
where $\lambda$ is a complex number.  For positive-frequency solutions
of the massless scalar wave equation, this inner product is
positive-definite [11].  By a suitable change of basis on each of
$\Sigma_I$ and $\Sigma_F$ independently, one can normalise the
$\{f_{\omega\ell m}\}$ and the $\{p_{\omega \ell m}\}$ such that 
$$\eqalignno{(f_{\omega\ell m}{\,},{\,}f_{\omega'\ell' m'}{\,}){\;}{\,}&
={\;}{\,}(p_{\omega\ell m}{\,},{\,}p_{\omega'\ell' m'}){\;}{\,} 
={\;}{\,}\delta_{\ell\ell'}{\;}\delta_{mm'}{\;}
\delta(\omega{\,},\omega'){\quad},&(2.11)\cr
(f_{\omega\ell m}{\,},{\,}f^{*}_{\omega'\ell' m'}){\;}{\,}&
={\;}{\,}(p_{\omega\ell m}{\,},{\,}p^{*}_{\omega'\ell' m'}){\;}{\,} 
={\;}{\,}0{\quad},&(2.12)\cr
(f^{*}_{\omega\ell m}{\,},{\,}f^{*}_{\omega'\ell' m'}){\;}{\,}&
={\;}{\,}(p^{*}_{\omega\ell m}{\,},{\,}p^{*}_{\omega'\ell' m'}){\;}{\,} 
={\;}{\,}-{\;}\delta_{\ell\ell'}{\;}\delta_{mm'}{\;}
\delta(\omega{\,},\omega'){\quad}.&(2.13)\cr}$$
\smallskip
\indent
Since the $\{f_{\omega'\ell m}\}$ form a complete orthonormal set on
$\Sigma_I{\,}$, one may expand out a typical basis function (solution)
${\,}p_{\omega\ell m}$ on the surface $\Sigma_I$ in the form
$$p_{\omega\ell m}{\;}{\,} 
={\;}{\,}\int^{\infty}_{0}{\;}d\omega'{\;}
\Bigl(\alpha_{\omega'\omega\ell m}{\;}f_{\omega'\ell m}{\;}
+{\;}\beta_{\omega'\omega\ell m}{\;}f^{*}_{\omega'\ell,-m}\Bigr){\quad},
\eqno(2.14)$$
\noindent
where the sets of complex numbers $\{\alpha_{\omega'\omega\ell m}\}$
and $\{\beta_{\omega'\omega\ell m}\}$ give the Bogoliubov coefficients
[11].  Here, from the definition of spherical harmonics
$Y_{\ell m}(\Omega)$ in [17], one has  
$Y_{\ell m}^{*}=(-1)^{m}{\;}Y_{\ell,-m}{\,}$.  Of course, the (approximate)
spherical symmetry of the background implies that solutions with the
same $\ell$ and ${\mid}m{\mid}$ are connected in Eq.(2.14).  Further,
spherical symmetry implies that the Bogoliubov coefficients are
independent of $m{\,}$, and we shall henceforth ignore this index.  
In the gravitational-collapse problem, the time-dependence of the collapse
geometry gives ${\,}f_{\omega\ell m}\neq{\,}p_{\omega\ell m}{\,}$; 
thus, the coefficients $\beta_{\omega'\omega\ell}$ are non-zero and
there is mixing between positive-and negative-frequency solutions.\par
\smallskip
\indent
The Bogoliubov coefficients may be expressed in terms of inner products as
$$\eqalignno{\alpha_{\omega'\omega\ell}{\;}{\,}&
={\;}{\,}(p_{\omega\ell m}{\,},{\,}f_{\omega'\ell m}{\,}){\quad},&(2.15)\cr
\beta_{\omega'\omega\ell}{\;}{\,}&
={\;}{\,}-{\;}(p_{\omega\ell m}{\,},{\,}
f^{*}_{\omega'\ell,-m}){\quad}.&(2.16)\cr}$$
\noindent
Given these relations, one may invert Eq.(2.14) by expanding out
$f_{\omega'\ell m}$ in terms of the $\{p_{\omega\ell m}\}$ and
$\{p^{*}_{\omega\ell,-m}\}$ basis on $\Sigma_F{\,}$, as
$$f_{\omega'\ell m}{\;}{\,} 
={\;}{\,}\int^{\infty}_{0}{\;}d\omega{\;}{\,}
\bigl(\alpha^{*}_{\omega'\omega\ell}{\;}p_{\omega\ell m}{\;}
-{\;}\beta_{\omega'\omega\ell}{\;}
p^{*}_{\omega\ell,-m}\bigr){\quad}.\eqno(2.17)$$
\noindent
One can similarly relate the 'Fourier amplitudes' 
$\{b_{\omega\ell m}\}$ and $\{c_{\omega'\ell m}\}$ above:
$$\eqalignno{b_{\omega\ell m}{\;}{\,}&
={\;}{\,}\int^{\infty}_{0}{\,}d\omega'{\;}{\,}
\bigl(\alpha^{*}_{\omega\omega\ell}{\;}c_{\omega'\ell m}{\;}
-{\;}\beta^{*}_{\omega'\omega\ell}{\;}c^{*}_{\omega'\ell,-m}\bigr)
{\quad},&(2.18)\cr
c_{\omega'\ell m}{\;}{\,}&
={\;}{\,}\int^{\infty}_{0}{\;}d\omega{\;}{\,}
\bigl(\alpha_{\omega'\omega\ell}{\;}b_{\omega\ell m}{\;}
+{\;}\beta^{*}_{\omega'\omega\ell}{\;}b^{*}_{\omega\ell,-m}\bigr)
{\quad}.&(2.19)\cr}$$
\noindent
Further, on substituting Eqs.(2.14,17) into Eqs.(2.11,13), one finds 
that the Bogoliubov coefficients must obey the quadratic relations
$$\eqalignno{&\int^{\infty}_{0}{\;}
d\omega{\;}{\,}\bigl(\alpha_{\omega''\omega\ell}{\;}
\alpha^{*}_{\omega'\omega\ell}{\;}
-{\;}\beta_{\omega'\omega\ell}{\;}\beta^{*}_{\omega''\omega\ell}){\;}{\,} 
={\;}{\,}\delta(\omega',{\,}\omega''){\quad},&(2.20)\cr
&\int^{\infty}_{0}{\;}d\omega{\;}{\,}
\bigl(\alpha^{*}_{\omega'\omega\ell}{\;}\beta_{\omega''\omega\ell}{\;}
-{\;}\beta_{\omega'\omega\ell}{\;}
\alpha^{*}_{\omega''\omega\ell}\bigr){\;}{\,} 
={\;}{\,}0{\quad},&(2.21)\cr
&\int^{\infty}_{0}{\;}d\omega''{\;}{\,}
(\alpha_{\omega''\omega\ell}{\;}\beta_{\omega''\omega'\ell}{\;} 
-{\;}\beta_{\omega''\omega\ell}{\;}\alpha_{\omega''\omega'\ell}){\;}{\,} 
={\;}{\,}0{\quad}.&(2.22)\cr}$$
\noindent
These equations (2.20-22), which naturally express the property that the
Bogoliubov coefficients are the matrix components for a map relating
orthonormal bases on $\Sigma_I$ and $\Sigma_F{\,}$, are discussed
further in [11].\par
\smallskip
\indent
This formalism, involving the use of Bogoliubov coefficients as in the
treatment of black-hole radiance and other consequences of quantum
field theory in curved space-time [9-12], can now be connected with
the alternative formalism set out in Secs.2-4 of [3].  Consider again 
the (nearly-) Lorentzian-signature case, where $T$ denotes the 
proper-time interval between the initial and final hypersurfaces.
Making the mode decomposition with respect to 'Lorentzian coordinates'
$(t,r,\theta,\phi)$:
$$\phi^{(1)}(t,r,\theta,\phi){\;}{\,} 
={\;}{\,}{1\over r}{\;}\sum^{\infty}_{\ell=0}{\;}\sum^{\ell}_{m=-\ell}
{\;}Y_{\ell m}(\Omega){\;}R_{\ell m}(t,r){\quad},\eqno(2.23)$$
\noindent
one arrives at the $(\ell{\,},m)$ mode equation:
$$\Bigl(e^{(b-a)/2}{\,}\partial_{r}\Bigr)^{2}R_{\ell m}
-{\;}\bigl(\partial_t\bigr)^{2}R_{\ell m}
-{\,}{1\over 2}\Bigl(\partial_{t}\bigl(a-b\bigr)\Bigr)
\bigl(\partial_{t}R_{\ell m}\bigr)
-{\,}V_{\ell}(t,r){\,}R_{\ell m}{\;}{\,}={\;}{\,}0{\quad}.\eqno
(2.24)$$
\noindent
Here,
$$V_{\ell}(t,r){\;}{\,} 
={\;}{\,}{{e^{b(t,r)}}\over{r^2}}{\;}\biggl({\,}\ell(\ell +1){\,}
+{\,}{{2m(t,r)}\over{r}}\biggr){\quad},\eqno(2.25)$$
\noindent
where $m(t,r)$ is defined in Eq.(2.2).  For adiabatic modes (that is, 
for frequencies $k$ of oscillation which are rapid compared to the
time rate of change of the background geometry), one has
approximately-separable solutions of the mode equation (2.24), of the
form
$$R_{\ell m}(t,r){\;}{\,} 
\sim{\;}{\,}\exp(ikt){\;}{\,}\xi_{k\ell m}(t,r){\quad},\eqno(2.26)$$
\noindent
where ${\,}\xi_{k\ell m}(t,r)$ varies 'slowly' with respect to
$t{\,}$. The definition of  ${\,}\xi_{k\ell m}(t,r)$ is tied down
completely at spatial infinity $(r\rightarrow\infty)$, where 
$R_{\ell m}(t,r)$ is required to reduce to a flat space-time separated
solution, for which ${\,}\xi_{k\ell m}(t,r)$ loses its
$t$-dependence.  By completeness, the general solution
$\phi^{(1)}(t,r,\theta,\phi)$ of the linear wave equation (2.3) may be
written as in Eq.(2.18) of [3]:
$$\phi^{(1)}{\;}{\,}
={\;}{\,}{1\over{r}}{\;}\sum^{\infty}_{\ell =0}{\;}
\sum^{\ell}_{m=-\ell}{\;}\int^{\infty}_{-\infty}{\,}dk{\;}{\,}
a_{k\ell m}{\;}\xi_{k\ell m}(t,r){\;}{{\sin(kt)}\over{\sin(kT)}}{\;} 
Y_{\ell m}(\Omega){\quad}.\eqno(2.27)$$ 
\noindent
where the $\{a_{k\ell m}\}$ are real coefficients.  Here, for
simplicity, we are taking, as in [2,3], boundary conditions on
$\phi^{(1)}$ for which $\phi^{(1)}{\mid}_{\Sigma_{I}}=0$ but
$\phi^{(1)}{\mid}_{\Sigma_{F}}\neq 0{\,}$.  Physically, if the
gravitational initial data $h_{ij}{\mid}_{\Sigma{I}}$ were also taken 
to be spherically symmetric, this would correspond to scalar particle 
production resulting from isotropic collapse data.  At very late
times, near the final hypersurface $\Sigma_F{\,}$, the geometry also 
varies extremely slowly with respect to time, and the 'adiabatic
separation functions' ${\,}\xi_{k\ell m}(t,r)$ again reduce to functions
of $r$ only, namely $\xi_{k\ell m}(r)$.  These obey the mode equation 
[1-3]
$$e^{(b-a)/2}{\;}{\partial\over{\partial r}}{\;}
\biggl(e^{(b-a)/2}{\;}{{\partial\xi_{k\ell}}\over{\partial r}}\biggr){\;} 
+{\;}\bigl(k^{2}-V_{\ell}\bigr){\;}\xi_{k\ell}{\;}{\,}
={\;}{\,}0{\quad},\eqno(2.28)$$
\noindent
with boundary conditions described below.  The spherical symmetry of
the background implies that ${\,}\xi_{k\ell m}(r)$ is independent of the
quantum number $m{\,}$.  Thus,
$$\xi_{k\ell m}(r){\;}{\,}={\;}{\,}\xi_{k\ell}(r){\quad}.\eqno(2.29)$$
\smallskip
\indent
Evaluating Eq.(2.27) at the final boundary $\Sigma_F$, one finds that
$$\phi^{(1)}(x)\Bigl\arrowvert_{\Sigma_F}{\;}{\,} 
={\;}{\,}{{1}\over{r}}{\;}\sum^{\infty}_{\ell =0}{\;}
\sum^{\ell}_{m=-\ell}{\;}\int^{\infty}_{-\infty}{\;}dk{\;}{\,}
a_{k\ell m}{\;}\xi_{k\ell}(r){\;}Y_{\ell m}(\Omega){\quad}.\eqno(2.30)$$
\noindent
This relation can be inverted (see Sec.3 of [3]), given the
normalisation of the $\{\xi_{k\ell}(r)\}$, to find the coefficients
$a_{k\ell m}$ in terms of the final data $\phi^{(1)}(x)$.  Indeed, the
$a_{k\ell m}$ characterise the final scalar data.  Since the mode 
functions $\{\xi_{k\ell}(r)\}$ on $\Sigma_F$ are real, one can rewrite 
this in the form
$$\phi^{(1)}(x)\Bigl\arrowvert_{\Sigma_F}{\;}{\,}
={\;}{\,}{{1}\over{r}}{\;}\sum^{\infty}_{\ell =0}{\;}
\sum^{\ell}_{m =-\ell}{\;}\int^{\infty}_{0}{\;}dk{\;}{\,}(a_{k\ell m}{\,}
+{\,}a_{-k\ell m}){\;}\xi_{k\ell}(r){\;}Y_{\ell m}(\Omega){\quad},
\eqno(2.31)$$
involving only positive $k$-values.  Here, the geometry in this 
space-time region is expected to be approximated very accurately by a 
Vaidya metric [13], corresponding to a smoothed-out luminosity in the 
radiation, which varies only slowly with time.  Such a metric can be
put in the diagonal form (2.1), with [13]:
$$e^{-a}{\;}{\,}
={\;}{\,}1{\,}-{\,}{{2m(t,r)}\over{r}}{\quad};{\qquad}{\quad}
e^{b}{\;}{\,}
={\;}{\,}\biggl({{\dot m}\over{f(m)}}\biggr)^{2}{\,}e^{-a}{\quad}.
\eqno(2.32)$$
Here, $m(t,r)$ is a slowly-varying function, with 
${\dot m}=(\partial m/\partial t)$, and the function $f(m)$ depends on 
the details of the radiation.\par
\smallskip
\indent
The 'left' boundary condition (3.1) of [3] reads
$$\xi_{k\ell}{\;}{\,}={\;}{\,}0{\quad}\eqno(2.33)$$
\noindent
at $r=0{\,}$.  The 'right' boundary condition (3.3) of [3] requires that:
$$\xi_{k\ell}(r){\;}{\,}
\sim{\;}{\,}\Bigl(z_{k\ell}{\;}\exp(ikr^{*}_{s}){\;}
+{\;}z^{*}_{k\ell}{\;}\exp(-ikr^{*}_{s})\Bigr){\quad},\eqno(2.34)$$
\noindent
as $r\rightarrow\infty{\,}$.  A generalisation $r^{*}$ of the standard
Regge-Wheeler coordinate $r_{s}^*$ for the Schwarzschild geometry [18]
may
be defined, by 
$${{\partial}\over{\partial r^{*}}}{\;}{\,}
={\;}{\,}e^{(b-a)/2}{\;}{{\partial}\over{\partial r}}{\quad}.
\eqno(2.35)$$
\noindent
Here, for each $(k,\ell),{\;}{\,}z_{k\ell}$ is a dimensionless complex
coefficient.  In our space-time, for sufficiently large $r{\,}$, a
coordinate $r^{*}_s$ may also be defined by 
$$r^{*}_{s}{\;}{\,} 
={\;}{\,}r{\,}+{\,}2M{\,}\ln\Bigl(\bigl(r/2M\bigr)-1\Bigr){\quad},
\eqno(2.36)$$
\noindent
where $M$ is the total ADM (Arnowitt-Deser-Misner) mass [18,19].  For
very large $r{\,}$, ${\,}r^*$ and $r_{s}^*$ are asymptotically equal.\par
\smallskip
\indent
In the nearly-Lorentzian case, the late-time behaviour of quantum 
amplitudes will (as in Sections 3,4 of [3]) be dominated by the
eigen-frequencies $k_n$ defined in Eq.(4.1) of [3]:
$$k_{n}{\;}{\,}={\;}{\,}{{n\pi}\over{T}}{\quad},\eqno(2.37)$$
\noindent 
$(n = 1,2,3,\ldots{\;}).{\,}$  Accordingly, let us discretise the frequency
integral in Eq.(2.31) above.  Define real functions $\{f_{n\ell}(r)\}$,
for $n=1,2,\ldots{\;};{\;}{\,}\ell =0,1,2,\ldots{\;}$, such that
$$\xi_{n\ell}(r){\;}{\,} 
={\;}{\,}\Bigl(z_{n\ell}{\;}e^{ik_{n}r^{*}}{\,}
+{\;}z^{*}_{n\ell}{\;}e^{-ik_{n}r^{*}}\Bigr){\;}f_{n\ell}(r){\quad}, 
\eqno(2.38)$$
\noindent
where ${\,}f_{n\ell}(r)\rightarrow 1{\,}$ as ${\,}r\rightarrow\infty{\;}$,
${\,}f_{n\ell}(r)\rightarrow 0{\,}$ as ${\,}r\rightarrow 0{\;}$, 
and ${\;}r^{*}{\;}\sim{\;}r^{*}_{s}{\;}$ as 
${\,}r\rightarrow\infty{\,}$.\par
\smallskip
\indent
The discrete description of Eq.(2.31) above is appropriate for the final
data ${\,}\phi^{(1)}(x){\mid}_{\Sigma_F}{\,}$, but we further need a 
discrete description of the space-time behaviour of
${\,}\phi^{(1)}(x)$.  For this, define ${\,}\Delta k_{n}=\pi/T{\,}$, 
and then (at least in a neighbourhood of $\Sigma_F$) define complex 
coefficients $A_{n\ell m}{\,}$, complex numbers $\hat z_{n\ell}$ and 
further real functions $\{g_{n\ell}(r)\}$ such that 
$$\phi^{(1)}(x){\;}{\,}
={\;}{\,}{1 \over r}{\;}\sum_{\ell mn}{\;}\Delta k_{n}{\;}A_{n\ell m}{\;}
\Bigl({\hat z}_{n\ell}{\;}e^{-ik_{n}(t-r^{*})}
+{\hat z}_{n\ell}{\;}e^{ik_{n}(t-r^{*})}{\,}\Bigr){\;}g_{n\ell}(r){\;}
Y_{\ell m}(\Omega){\quad},\eqno(2.39)$$ 
\noindent
where the $\{g_{n\ell}(r)\}$ obey the same boundary conditions as the
$\{f_{n\ell}(r)\}{\,}$.  Eq.(2.39) agrees with the discretised version 
of Eq.(2.31), provided that 
$$\eqalign{g_{n\ell}(r){\;}{\,}&
={\;}{\,}f_{n\ell}(r){\quad},{\qquad}{\quad}z_{n\ell}{\;}{\,} 
={\;}{\,}{\hat z}_{n\ell}{\;}{\,}e^{-ik_{n}{\mid}T{\mid}}{\quad},\cr
A_{n\ell m}{\;}{\,}&
={\;}{\,}a_{n\ell m}{\,}+{\,}a_{-n\ell m}{\quad}.\cr}\eqno(2.40)$$
\indent
In order to make a comparison with the Bogoliubov description, consider
now the discretised version of Eq.(2.5), appropriate to the final
surface $\Sigma_F{\,}$, taking
$$p_{n\ell m}(x){\;}{\,} 
={\;}{\,}N(k_{n}){\;}{{p_{n\ell}(r)}\over{r}}{\;}e^{-ik_{n}(t-r^{*})}{\;} 
Y_{\ell m}(\Omega){\quad}.\eqno(2.41)$$
\noindent
Given the boundary condition of regularity at ${\,}r=0{\,}$, Eq.(2.7) shows
that the normalisation factor is
$$N(k_{n}){\;}{\,}={\;}{\,}(2\pi k_{n})^{-{1/2}}{\quad}.\eqno(2.42)$$
\noindent
Then, comparing the discretised form of Eq.(2.31) with Eq.(2.39), we find
$$p_{n\ell}(r){\;}{\,}={\;}{\,}f_{n\ell}(r){\quad},\eqno(2.43)$$
$$b_{n\ell m}{\;}{\,} 
={\;}{\,}{{(a_{n\ell m}+{\,}a_{-n\ell m}){\,}
{\hat z}_{n\ell}}\over{N(k_{n})}}{\quad}.\eqno(2.44)$$
\noindent
In calculating particle emission rates in the following Section 3, we
shall need
$${\mid}b_{n\ell m}{\mid}^{2}{\;}{\,} 
={\;}{\,}2\pi{\,}k_{n}{\,}{\mid}z_{n\ell}{\mid}^{2}{\;}
{\mid}a_{n\ell m}+a_{-n\ell m}{\mid}^{2}{\;}{\,} 
={\;}{\,}2{\,}f_{\ell m}(k_{n}){\quad},\eqno(2.45)$$
\noindent 
where we have used the fall-off property as 
${\,}{\mid}k{\mid}\rightarrow\infty{\,}$:  
$${\mid}f_{\ell m}(k){\mid}{\;}{\,}
\sim{\;}{\,}{\mid}k{\mid}^{-3}\eqno(2.46)$$
\noindent
of Eq.(4.6) of [3].  For later use, we also record an expression for 
${\mid}b_{\omega\ell m}{\mid}^2$ in terms of Bogoliubov coefficients, 
where Eq.(2.18) has been used. Including an implicit summation over
the index $m{\,}$, we find
$$\eqalign{{\mid}b_{\omega\ell m}{\mid}^{2}{\;}{\,}&
={\;}{\,}\int^{\infty}_{0}{\,}d\omega'{\,}\int^{\infty}_{0}{\,}
d\omega''{\;}{\,}\biggl(\alpha^{*}_{\omega'\omega\ell}{\;}
\alpha_{\omega''\omega\ell}{\;}    
+{\;}\beta^{*}_{\omega''\omega\ell}{\;}\beta_{\omega'\omega\ell}\biggr){\;}
c_{\omega'\ell m}{\;}c^{*}_{\omega''\ell m}\cr
&-2\int^{\infty}_{0}{\,}d\omega'{\,}\int^{\infty}_{0}{\,}d\omega''{\;}{\,}
{\rm Re}\biggl(\beta_{\omega''\omega\ell}{\;}
\alpha^{*}_{\omega'\omega\ell}{\;} 
c_{\omega'\ell m}{\;}c_{\omega''\ell,-m}\biggr){\quad}.\cr}\eqno(2.47)$$
\noindent
{\bf 3. Particle emission rates}
\medskip
\indent
As remarked in Sec.5 of [2], the total energy of the final
Einstein/scalar field configuration must equal that of the original 
configuration, namely $M{\,}$, the true ADM mass of the 'space-time'.  
Recall that we are here considering gravitational and massless-scalar 
perturbations about a given spherically-symmetric background solution, 
of the form  
$g_{\mu\nu} 
=\gamma_{\mu\nu}+\epsilon{\;}h^{(1)}_{\mu\nu}{\,}
+\epsilon^{2}{\;}h^{(2)}_{\mu\nu}+\ldots{\;},
{\quad}\phi
=\Phi+{\,}\epsilon{\,}\phi^{(1)}
+{\,}\epsilon^{2}{\,}\phi^{(2)}+\ldots{\;}$.  
The 'second-variation' $O(\epsilon^{2})$ contribution to $M$ (that is,
the leading contribution beyond the background contribution) on 
$\Sigma_F$ is
$$E^{(2)}{\;}{\,} 
={\;}{\,}\int_{\Sigma_F}{\,}d^{3}x{\;}(-\gamma)^{1\over 2}{\;}{\cal H}
{\quad},\eqno(3.1)$$
\noindent
where ${\cal H}=e^{-b}{\,}T^{(2)}_{tt}$ and where $T^{(2)}_{\mu\nu}$ 
is the perturbed energy-momentum tensor, given in Eq.(3.28) of [2], 
which is of quadratic order in the scalar-field perturbations, and, 
if evaluated at a late time, has no contribution from the background 
scalar field $\Phi(t,r)$, in view of the final boundary conditions
that the gravitational data are spherically symmetric on $\Sigma_F{\,}$.  
From the wave equation (2.3), integrating by parts, using the boundary 
conditions of regularity and the asymptotic behaviour  
${\,}r^{2}{\;}\phi^{(1)}{\,}\partial_{r}\phi^{(1)}{\,}={\,}O(r^{-1}){\,}$ 
as $r\rightarrow\infty{\,}$, we find
$$E^{(2)}{\;}{\,} 
={\;}{\,}{1 \over 2}\int d\Omega\int_{\Sigma_F}{\,}dr{\;}
r^{2}{\;}e^{(a-b)/2}{\;}
\biggl[\dot\phi^{(1)2}{\,}-{\,}\phi^{(1)}{\;}\ddot\phi^{(1)}{\,} 
-{\,}\Bigl({1\over 2}\Bigl){\,}(\dot a-\dot b){\;}
\phi^{(1)}{\;}\dot\phi^{(1)}{\,}\biggr]{\quad}.\eqno(3.2)$$
\noindent
We may neglect the terms in Eq.(3.2) with integrand proportional to
$(\dot a -\dot b)$, compared with the 
$\bigl[{\,}\dot\phi^{(1)2}-\phi^{(1)}\ddot\phi^{(1)}{\,}\bigr]$ terms,
provided that the adiabatic approximation of Sec.2 of [3] holds; that
is, that typical perturbative frequencies ${\,}\omega{\,}$ obey 
${\,}\omega\gg{1 \over 2}{\mid}\dot a-\dot b{\mid}{\,}$.\par
\smallskip
\indent
In the adiabatic case, $E^{(2)}$ decomposes into a 'sum' over
frequencies.  On substituting the representation (2.5) into Eq.(3.2) and
using Eq.(2.41), we obtain
$$E^{(2)}{\;}{\,} 
={\;}{\,}\sum_{\ell m}{\;}\int^{\infty}_{0}{\,}d\omega{\;}{\,}\omega{\;}
{\mid}b_{\omega\ell m}{\mid}^{2}{\quad}.\eqno(3.3)$$
\noindent
This can also be re-written in the 'harmonic-oscillator coordinates'
$\{c_{\omega\ell m}\}$, on using Eq.(2.47).\par
\smallskip
\indent
One can read off the final particle-number spectrum from the total
energy by writing
$$E^{(2)}{\;}{\,} 
={\;}{\,}\sum_{\ell m}{\;}\int^{\infty}_{0}{\,}d\omega{\;}{\,}
\omega{\;}{{dN_{\omega\ell m}}\over{d\omega}}{\quad}.\eqno(3.4)$$
\noindent
On using Eq.(2.45), one obtains
$${{dN_{\omega\ell m}}\over{d\omega}}{\;}{\,} 
={\;}{\,}{\mid}b_{\omega\ell m}{\mid}^{2}{\;}{\,}
={\;}{\,}2{\,}f_{\ell m}(\omega){\quad},\eqno(3.5)$$
\noindent
where $f_{\ell m}(\omega)$ is defined through Eqs.(2.41-43).  Let 
$\Delta G_{\omega}$  be the number of states or phase-space cells in mode
$\omega{\,}$, and let ${N'}_{\omega\ell m}$ be the number of (scalar)
particles in these states. Then
$${{{N'}_{\omega\ell m}}\over{\Delta G_{\omega}}}{\;}{\,} 
={\;}{\,}<n_{\omega\ell m}>\eqno(3.6)$$ 
\noindent
is the mean number of particles in each of the (quantum) states in
mode $(\omega\ell m)$.  Since the number of states emitted in the
frequency interval $(\omega{\,},{\,}\omega +d\omega)$ is
$$\Delta G_{\omega}{\;}{\,} 
={\;}{\,}{{{\mid}T{\mid}}\over{\pi}}{\;}d\omega{\quad},\eqno(3.7)$$
\noindent
the number of particles in the range $(\omega{\,},{\,}\omega +d\omega)$ is
$$d{N'}_{\omega\ell m}{\;}{\,} 
={\;}{\,}{{{\mid}T{\mid}}\over{\pi}}{\;}<n_{\omega\ell m}>{\;}d\omega{\quad}.
\eqno(3.8)$$
\noindent
In the continuum limit,
$$d{{\dot N}'}_{\omega\ell m}{\;}{\,}
={\;}{\,}{{<n_{\omega\ell m}>}\over{\pi}}{\;}d\omega\eqno(3.9)$$
\noindent
is the total emission rate in the range $(\omega{\,},{\,}\omega +d\omega)$,
where ${\,}{{\dot N}'}_{\omega\ell m}{\,}$ is the number of particles 
emitted over the duration ${\mid}T{\mid}{\,}$.  (The factor $\pi{\,}$, 
rather than the usual factor $2\pi{\,}$, occurs in the denominator in 
Eq.(3.9) due to the initial time being $t = 0$ rather than 
${\,}t=-{\mid}T{\mid}{\,}$.)  Alternatively, Eq.(3.9) just gives the 
number of particles passing out through the surface of a sphere
centred on the collapsing 'star' or black hole.\par
\smallskip
\indent
In the case (as here) that the particles emitted do not decay further,
one identifies $<n_{\omega\ell m}>{\,}$ as [11]
$$<n_{\omega\ell m}>{\;}{\,} 
={\;}{\,}{\mid}\beta_{\omega\ell}{\mid}^{2}{\quad},\eqno(3.10)$$ 
\noindent
where ${\mid}\beta_{\omega\ell}{\mid}^{2}$ is defined by
$$\int^{\infty}_{0}d\omega'{\;}{\,}
\beta_{\omega'\omega\ell}{\;}\beta^{*}_{\omega'\omega''\ell}{\;}{\,} 
={\;}{\,}{\mid}\beta_{\omega\ell}{\mid}^{2}{\;}\delta(\omega,\omega'')
{\quad}.\eqno(3.11)$$
\noindent
Then the total number spectrum, added up over the black-hole life-time, 
is
$$\eqalign{{{d{N'}_{\omega}}\over{d\omega}}{\;}{\,}&
={\;}{\,}{{1}\over{\pi}}{\;}\sum_{s\ell m}{\,}N_{s\omega\ell m}{\;}
\int^{t_F}_{t_I} dt{\;}{\,}{\mid}\beta_{s\omega\ell m}{\mid}^{2}\cr
&={\;}{\,}{{1}\over{\pi}}{\;}\sum_{s\ell m}{\,}N_{s\omega\ell m}{\;} 
\int^{E}_{0}{\;}dM{\;}{\,}\biggl({{-dM}\over{dt}}\biggr)^{-1}{\;}
{\mid}\beta_{s\omega\ell m}{\mid}^{2}{\quad},\cr}\eqno(3.12)$$
\noindent
where $N_{s\omega\ell m}$ counts the number of states with spin $s$
and quantum numbers $(\omega\ell m)$.  In the original calculation of
Bogoliubov coefficients and of probabilities for particle emission by
a non-rotating black hole [10], it was found (neglecting the effects of 
back-reaction) that
$${\mid}\beta_{s\omega\ell m}{\mid}^{2}{\;}{\,} 
={\;}{\,}\Gamma_{s\omega\ell m}(\tilde m){\;}
\Bigl(e^{4\pi{\tilde m}}-(-1)^{2s}\Bigr)^{-1}{\quad},\eqno(3.13)$$
\noindent
where $\Gamma_{s\omega\ell m}({\tilde m})$ is the transmission
probability over the centrifugal barrier [20] and ${\tilde m}=2M\omega$
is dimensionless.  This calculation, of course, referred to the case 
where one did not have a final surface $\Sigma_F$ of topology 
${\Bbb R}^{3}$.  But, because of the very-high-frequency (adiabatic) 
method through which this expression for 
${\mid}\beta_{s\omega \ell m}{\mid}^{2}$ was calculated, it should 
still be valid (up to tiny corrections) in our present ${\Bbb R}^{3}$ 
case.\par
\smallskip
\indent
The semi-classical expression for the rate of mass loss is [21]
$${{dM}\over{dt}}{\;}{\,} 
={\;}{\,}-{\;}{{\alpha_{0}(M)}\over{M^{2}}}{\quad},\eqno(3.14)$$
\noindent
where $\alpha_{0}(M)\simeq{\rm constant}$ for massless or
ultra-relativistic particles, and $M = M(M_{I}, t)$ is the mass to
which a black hole of initial mass $M_I$ has been reduced after time
$t{\,}$.  Turning again to the total number spectrum over the black-hole
life-time, given by ${N'}_{\omega}$ as in Eq.(3.12), note that, for
massless particles, the Bogoliubov coefficients
${\mid}\beta_{s\omega\ell m}{\mid}^{2}$ depend on $\omega$ only through 
$\tilde m$ (on dimensional grounds).  Hence,
$${{d{N'}_{\omega}}\over {d\omega}}{\;}{\,} 
={\;}{\,}{{1}\over{8\pi\alpha_{0}{\,}\omega^{3}}}{\;}
\sum_{s\ell m}{\;}\int^{2{M_I}\omega}_{0}{\;} d{\tilde m}{\;}{\,}
{\tilde m}^{2}{\;}{\mid}\beta_{s\omega\ell m}(\tilde m){\mid}^{2}
{\quad}.\eqno(3.15)$$
\noindent
In particular, in considering the high-frequency limit 
${\,}\omega M_{I}\gg 1{\,}$ of this expression, we may replace the 
upper limit in the $\tilde m$-integral by infinity.  From this, one 
finds that the time-integrated energy distribution of all the massless 
particles radiated by the black hole (of initial mass $M_I$) is
$${{d{N'}_{\omega}}\over{d\omega}}{\;}{\,} 
\sim{\;}{\,}c_{1}{\;}\omega^{-3}{\quad},\eqno(3.16)$$
\noindent
for ${\,}\omega\gg(M_I)^{-1}{\,}$, where $c_1$ is a real number, so 
giving the form of the high-energy tail of the spectrum.  However, most
of the particles produced during the quasi-stationary regime of the
radiating black hole, before its expected explosive phase with
$M\rightarrow 0{\,}$, are at a temperature corresponding to the initial
mass of the hole $\Bigl(\omega\sim (M_{I})^{-1}\Bigr)$, where the energy
distribution peaks.\par
\smallskip
\indent
How is the above analysis related to our theory which includes
back-reaction?  One might expect that the high-energy behaviour of the
$\{a_{\omega\ell m}\}$ 'coordinates', which describe the perturbed
scalar field $\phi^{(1)}$ on the final surface $\Sigma_F{\,}$, would be
related to the emission in the final moments of evaporation.  This can
be seen from Eq.(3.5), taking the frequencies 
$k_{n}=\sigma_{n}=n\pi/{\mid}T{\mid}{\,}$, without an explicit 
calculation of the Bogoliubov coefficients.  The high-energy behaviour 
of the function $f_{\ell m}(\omega)$, which is defined through
Eq.(4.3) of [3], was found in Eq.(4.6) of [3] from the requirement
that the contour at infinity in the integral (4.2) of [3] for the 
action should contribute zero.  For massless scalar particles, this 
requirement, expressed through Eq.(3.5), reads
$${{dN_{n\ell m}}\over{d\omega}}{\;}{\,} 
\propto{\;}{\,}(\sigma_{n})^{-3}{\quad},\eqno(3.17)$$
for ${\,}\sigma_{n}\gg (M_I)^{-1}{\,}$, in agreement with the 
high-energy behaviour in Eq.(3.16).  Here, again, the requirement for 
a finite imaginary part for the classical Lorentzian action 
$S^{(2)}_{\rm class}$, corresponding to a non-trivial probability 
distribution over final configurations, is linked with the behaviour 
of the high-energy number spectrum; the latter is of course subject in
principle to observational test.  \par
\smallskip
\indent
It is well known that the thermal equilibrium between a black hole and
the exterior radiation is unstable, as the specific heat in the
canonical ensemble is negative [22].  The canonical ensemble
breaks down for black holes since the canonical partition function
diverges for all temperatures.  Rather than work at fixed
temperature, one must consider the micro-canonical ensemble, which is
tailored to configurations of fixed energy.  At the high-energy end of
the emission spectrum, therefore, when the black hole approaches the
Planck scale, the canonical distribution, as given by Eq.(3.13), must
be replaced by the micro-canonical distribution.  Naturally, the
mass-loss rates for the two ensembles differ considerably.  The
micro-canonical decay rate modifies the small-mass behaviour of the
black hole, where Eq.(3.14) breaks down.  For the low-frequency quanta
$(\omega\ll M)$ characteristic of the majority of the evaporation
process, the canonical and micro-canonical ensembles are almost
equivalent, and one obtains a Planck-like number spectrum Eq.(3.13)
[23].\par
\smallskip
\indent
Prior to the black hole's total evaporation, it is possible for the
black hole to emit a single high-energy quantum with energy comparable
to the initial black-hole mass. As ${\mid}\beta_{j}{\mid}^2$ is the
average number of quanta present, then in the high-energy tail of the
spectrum, one has ${\mid} \beta_{j}{\mid}^{2}\sim e^{-S}$, 
where $S$ is the Boltzmann entropy of the system, since there
is only one state out of a total of $e^{S}$ states for which all the
energy is concentrated into one quantum [24].
More precisely,
$${{{\mid}\beta_{j}{\mid}^{2}} \over{{\mid} \alpha_{j}{\mid}^{2}}}\;\;
\sim\;\; e^{-\Delta S_{BH}}\quad , \eqno(3.18)$$
where $\Delta S_{BH}$ is the difference in black-hole entropy before
and after emission.  This result is expected to be valid for a
general spherically-symmetric black hole.\par
\medskip
\noindent
{\bf 4. Probabilistic interpretation}
\medskip
\indent
When ${\mid}T{\mid}{\,}$ is large but finite, one finds from Eq.(4.16) 
of [3], that the density function
$$P\Bigl[\{a_{k\ell m}\};{\mid}T{\mid}\Bigr]{\;}{\,}
={\;}{\,}{\hat N}{\;}e^{-\delta{\mid}T{\mid}M_{I}}{\;}
\exp\Bigl(-{\,}2{\,}{\rm Im}{\;}S^{(2)}_{\rm class}
\bigl[\{a_{k\ell m}\};{\mid}T{\mid}\bigr]\Bigr){\quad},\eqno(4.1)$$
\noindent
where $\hat N$ is a suitable normalisation factor, describes the
conditional probability density over the final perturbative scalar
boundary data $\phi^{(1)}{\mid}_{\Sigma_F}{\,}$, the condition being 
that the perturbations obey the initial conditions
$\phi^{(1)}{\mid}_{\Sigma_I}{\,}=0$ of Eq.(2.7) of [3].  The limit 
$\delta\rightarrow 0$ should then be taken;  the only data at spatial
infinity itself consist of the proper time ${\mid}T{\mid}{\,}$. This
probability is, of course, the squared norm of a complex quantum
amplitude, in the context of [3].\par
\smallskip
\indent
Even though the probability distribution (4.1) arises from squaring a
quantum amplitude, one can still ask, given the thermal nature of
black-hole evaporation in the usual description, whether the
probability distribution (4.1) can be viewed in terms of the diagonal
components of some non-trivial 'density matrix' in a suitable basis
[9-12,32].  To evaluate more explicitly the probability distribution,
consider Eq.(2.36), using also Eqs.(2.20,45,47) and
Eq.(4.16) of [3].  One finds, in the continuum limit of large 
${\mid}T{\mid}$, taking the $\{a_{k\ell m}\}$ as the final 
'coordinate' variables, that:
$$\eqalign{P\Bigl[\{a_{\omega\ell m}\}\Bigr]{\;}{\,}&
={\;}{\,}{\hat N}{\;}\exp\biggl(-{\,}2\sum_{\ell m}{\;}
\int^{\infty}_{0}{\,}d\omega{\;}{\,}
{\mid}b_{\omega\ell m}{\mid}^{2}\biggr)\cr
&={\;}{\,}{\hat N}{\;}\exp\Biggl\{-{\,}2\sum_{\ell m}{\;}
\biggl[\int^{\infty}_{0}{\,}d\omega{\;}{\,}
{\mid}c_{\omega\ell m}{\mid}^{2}\cr
&+{\;}2\int^{\infty}_{0}{\,}d\omega'{\;}{\,}\int^{\infty}_{0}{\,}
d\omega''{\;}{\,}c_{\omega'\ell m}{\;}c^{*}_{\omega''\ell m}{\;}
\int^{\infty}_{0}{\;}d\omega{\;}{\,}
\beta^{*}_{\omega''\omega\ell}{\;}\beta_{\omega'\omega\ell}\cr
&-{\;}2\int^{\infty}_{0}{\,}d\omega'{\;}{\,}\int^{\infty}_{0}{\;}
d\omega''{\;}{\,}{\rm Re}\biggl(c_{\omega'\ell m}{\;}c_{\omega''\ell,-m}{\;}
\int^{\infty}_{0}{\;}d\omega{\;}{\,}\beta_{\omega''\omega\ell}{\;}
\alpha^{*}_{\omega'\omega\ell}{\,}\biggr)\biggr]\Biggr\}
{\quad}.\cr}\eqno(4.2)$$
\noindent
The first two terms inside the square bracket are positive-definite,
whereas the third is of indefinite sign.  The situation becomes much
clearer in the case that
$$\int^{\infty}_{0}{\,}d\omega{\;}{\,}\beta^{*}_{\omega''\omega\ell}{\;}
\beta_{\omega'\omega\ell}{\quad} 
={\quad}{\mid}\beta_{\omega'\ell}{\mid}^{2}{\;}{\,}
\delta(\omega',\omega''){\quad},\eqno(4.3)$$
$$\int^{\infty}_{0}{\;}d\omega{\;}{\,}\beta_{\omega''\omega\ell}{\;}
\alpha^{*}_{\omega'\omega\ell}{\;}{\,} 
={\;}{\,}0{\quad}.\eqno(4.4)$$
\noindent
This holds, in particular, for the steady-state Bogoliubov
coefficients in the calculation which neglects back-reaction on the
metric [10].  It remains to check whether this diagonal form persists
under adiabatic propagation through a slowly-varying potential.  From 
Eq.(2.20), one has
$${\mid}\alpha_{\omega'\ell}{\mid}^{2}{\,}
-{\,}{\mid}\beta_{\omega'\ell}{\mid}^{2}{\;}{\,}
={\;}{\,}1{\quad}.\eqno(4.5)$$
\noindent
Thence,
$$\eqalign{P&\Bigl[\{a_{\omega\ell m}\}\Bigr]{\;}{\,} 
={\;}{\,}P\Bigl[\{c_{\omega\ell m}\}\Bigr]\cr
&={\;}{\,}{\hat N}{\;}\exp\Biggl[-{\,}2{\;}
\sum_{\ell m}{\;}\int^{\infty}_{0}{\,}d\omega{\;}{\,}
{\mid}c_{\omega\ell m}{\mid}^{2}{\,}-{\,}4{\;}\sum_{\ell m}{\;}
\int^{\infty}_{0}{\;}d\omega{\;}{\,}{\mid}\beta_{\omega\ell}{\mid}^{2}{\;}
{\mid}c_{\omega\ell m}{\mid}^{2}\Biggr]\cr
&={\;}{\,}{\hat N}{\;}\prod_{n\ell m}{\;}
\exp\biggl[-{\,}2{\,}(\Delta\omega_{n}){\;}{\mid}c_{n\ell m}{\mid}^{2}{\,}
-{\;}4{\;}(\Delta\omega_{n}){\;}
{\mid}\beta_{n\ell}{\mid}^{2}{\;}{\mid}c_{n\ell m}{\mid}^{2}\biggr]
{\quad},\cr}\eqno(4.6)$$
\noindent
where ${\,}\Delta\omega_{n}{\,}={\,}\pi/{\mid}T{\mid}{\,}$.  The
product over ${\,}n\ell m$ tells us that the modes evolve
independently; such uncorrelated modes are a consequence of the
linearised theory.\par
\smallskip
\indent
A further argument for the diagonal form of the Bogoliubov
coefficients is through the tunnelling interpretation of black-hole
evaporation.  Particles are created in pairs just outside the future
horizon $\cal {H}^+$, with one member always falling into the
singularity whilst the other escapes to infinity, or is reflected back
down the hole.  By constructing a basis of scalar field modes which is
continuous across the future horizon, we automatically incorporate
both the positive- and negative-energy particles [25]. 
To achieve this in the Schwarzschild picture, however, one
must avoid the coordinate singularity at the future horizon by adding
a small imaginary part to the mass $M$.  Such a procedure is in
keeping with Feynman's complex-time technique -- which we expound in
this paper -- since the conjugacy between the intial mass $M_I$ and
asymptotic time $T$ implies that moving $T$ slightly into the lower
complex plane is equivalent to adding a small positive imaginary part
to $M_I$.  The Bogoliubov coefficients are just the weights of the
outgoing and ingoing particle components of the scalar field across
the future horizon. \par
\smallskip
\indent
Equation (4.6) expresses the probability distribution of final field
configurations in terms of amplitudes associated with the $\cal {S}^-$
Cauchy surface.  The nature of Eq.(4.6) suggests that there is a shift
in the width of the ground-state probability distribution.  That is,
$P[\{c_{n\ell m}\}]$ appears as an excited state relative to the
ground state probability density $P[\{b_{n \ell m}\}]$.  Having
averaged over $\{c_j\}$ amplitudes, the stability of the $\cal {S}^-$
state is guaranteed by the fact that $\,{\mid} \alpha{j}{\mid}^{2}
+{\mid}\beta_{j}{\mid}^{2}\,>\,0\,$ for all $j$.  However, the shift only
occurs for the infra-red or low-frequency components of the field, if
we assume that the Bogoliubov coefficients ${\mid} \beta_{j}{\mid}^{2}$
decay (exponentially) rapidly at high frequency and are finite for low
frequency.  This is just an expression of local (coordinate)
covariance, in that the $\Sigma_F$ and $\cal{S}^-$ representations are
equivalent for the ultra-violet properties of the theory.  An
infra-red ambiguity is manifest through the occurrence of the Hawking
effect with non-zero ${\mid}\beta_{j}{\mid}^2$.  Hence, the Hawking
effect is a consequence of the ambiguity in the choice of space-like
evolution hypersurfaces [26].\par
\smallskip
\noindent
{\bf 5. Conclusion}
\medskip
\indent
In this paper, we have made a connection between the present
boundary-value approach, as used in this study of quantum amplitudes
in black-hole evaporation, and the original approach by means of
Bogoliubov coefficients.  This connection is established through
consideration of the radial equation obeyed by the (adiabatic)
perturbations; in this paper, it is the spin-0 perturbations which are
studied.  The discussion in Sec.4 gives the probability
distribution for configurations of the perturbative scalar field, on a
final hypersurface $\Sigma_F$ at a very late time,
by taking the squared norms of amplitudes, as a product of independent
Gaussians.  Since our approach gives a final pure state, whereas the
traditional approach gives a non-trivial density matrix, 
one might expect some difference (not necessarily
great) between the probabilistic predictions of our boundary-value
approach and of the Bogoliubov-coefficient approach, in the case of
gravitational collapse to a black hole.  In the Appendix, we
consider a somewhat different question, and find a possible
interpretation of our probability
distribution, in terms of the Wigner quasi-probability distribution
for harmonic oscillators.\par
\smallskip
\indent
Further work in this area  has involved the study of 
approximately Vaidya-like metrics [13], as giving an accurate 
approximation to the gravitational field in the region of space-time 
containing the flux of outgoing black-hole radiation.  At the same 
time, we discuss the radiative spin-0 (scalar) and spin-2 (graviton) 
fields which both act as a source for the gravitational field and 
propagate within it.  This Vaidya-like description is essential in the 
treatment above of the adiabatic perturbation modes.  In later work,  
we shall give a further alternative description of the quantum states 
found in our boundary-value approach, namely, a description in terms
of coherent and squeezed states.  In [27,28] we repeat the spin-0
boundary-value quantum calculation of [3] (as used in the present
paper), but for other spins, including the fermionic 
spin-$1 \over 2$ (neutrino) case.\par
\medskip
\noindent
{\bf Appendix: Density-Matrix Interpretation and Wigner Distribution
Functions} 
\medskip
\indent
Our probability distribution $P[\{a_{\omega\ell m}\}] =P[\{c_{\omega
\ell m}\}]$ above naturally arises from the pure state that we have
discussed.  But we now check whether or not $P[\{c_{\omega \ell m}\}]$
can be expressed in terms of the diagonal elements of some other kind
of 'density-matrix distribution'.  One might expect that a probability
density for the final scalar field would incorporate not only the
randomness in the particle-emission process but also possible choices
for the boundary data.  In light of the loss of phase information in
Eq.(4.6) with respect to the $\{c_{\omega \ell m}\}$ amplitudes, a
diagonal density-matrix-like expression for $P[\{c_{\omega \ell m}\}]$
might now be possible.  Eq.(4.6) can be rewritten so as to make this clear,
when one employs
the generating function for the Laguerre polynomials 
$\{L_{k}(x)\}{\,}$ [29]:
$${{1}\over{(1-s)}}\exp\biggl[{{-xs}\over{(1-s)}}\biggr]{\;}{\,} 
={\;}{\,}\sum^{\infty}_{k=0}{\;}L_{k}(x){\,}s^{k},
{\qquad}{\mid}s{\mid}{\,}<{\,}1{\quad},\eqno(A1)$$ 
\noindent
where
$$L_{k}(x){\;}{\,} 
={\;}{\,}\sum^{k}_{n=0}{\;}{\,}{k\choose n}{\;}{\,}
{{(-x)^{n}}\over{n!}}{\quad}.\eqno(A2)$$
\noindent
Here, we set
$$s{\;}{\,} 
={\;}{\,}{\Bigl\arrowvert}
{{\beta_{n\ell}}\over{\alpha_{n\ell}}}{\Bigl\arrowvert}^{2}{\;}<{\;}1
\eqno(A3)$$
\noindent
and 
$$x{\;}{\,}={\;}{\,}2{\,}X_{n\ell m}{\;}{\,}
={\;}{\,}4{\,}(\Delta\omega_{n}){\;}{\mid}c_{n\ell m}{\mid}^{2}{\quad},
\eqno(A4)$$
\noindent
a dimensionless quantity.  Then, from Eqs.(4.6,A1), and writing 
$j=n\ell m$, we have
$$\eqalign{P\Bigl[\{c_{j}\}\Bigr]{\;}{\,}&
={\;}{\,}{\hat N}{\;}\prod_{j}{\;}\exp(-X_{j}){\;}
\exp\Bigl(-{\,}2{\;}{\mid}\beta_{j}{\mid}^{2}{\;}X_{j}\Bigr)\cr
&={\;}{\,}{\hat N}{\;}\prod_{j}{\;}\exp(-X_{j}){\;}
\sum^{\infty}_{k_{j}=0}{\;}P(k_{j}){\;}{\,}L_{k_j}(2X_{j})\cr
&={\;}{\,}{\hat N}{\;}\sum_{\{k_{j}\}}{\;}P\bigl(k_{j}\bigr){\;}
\prod_{j}{\;}\exp(-X_{j}){\;}{\,}L_{k_{j}}(2X_{j}){\quad},\cr}\eqno(A5)$$
\noindent
where $P(k_{j})$ is defined to be
$$P(k_{j}){\;}{\,} 
={\;}{\,}{{1}\over{{\mid}\alpha_{j}{\mid}^{2}}}{\;}
{\Bigl\arrowvert}{{\beta_{j}}\over{\alpha_{j}}}
{\Bigr\arrowvert}^{2k_{n\ell m}}{\quad},\eqno(A6)$$
\noindent   
namely the probability to observe the field in the state $k{\,}$, such
that from Eq.(2.20) one has [30] 
$$\sum^{\infty}_{k_{j}{\,}={\,}0}P(k_{j}){\;}{\,}
={\;}{\,}1{\quad}.\eqno(A7)$$
\noindent
Alternatively, $P(k_j)$ is the probability of finding $k$ particles
outgoing at ${\cal I}^{+}$ (future null infinity), in the mode
labelled by $n\ell m{\,}$.  The average number of particles outgoing at
${\cal I}^{+}$ per unit frequency around $\omega_n$ per unit time is
then $<k_j>={\,}{\mid}\beta_{j}{\mid}^2{\,}$, independently 
of $m{\,}$.  The factor
${{{\mid}{\beta_{j}}{\mid}^{2}}\over{{\mid}\alpha_{j}{\mid}^{2}}}$ effectively
controls the probability per unit time to emit particles in the
frequency range around $\omega_n$.  Each radiation state is uniquely
specified by the set of occupation numbers $\{k_{j}\}$ of the various
modes $\{n\ell m\}$ and occurs with probability $P(\{k_{j}\})$.\par
\smallskip
\indent
Eq.(A5) describes the probability distribution $P[\{c_{j}\}]$ for the
final scalar field, with the help of the Bose-Einstein distribution 
(A6), which describes the randomness in the particle emission
process.
Since, from Eq.(4.6), $P[\{c_{j}\}]$ depends only on the modulus
${\mid}c_j{\mid}$ of the complex number $c_j{\,}$, the sum in
Eq.(A5) is diagonal.  Of course, the explicit  expression for the 
'probability' (A5) for a given total mass $M_I$ is clearly
complicated, since, as in Eq.(3.13), the Bogoliubov coefficients 
${\mid}\beta_{j}{\mid}^2$ depend on the transmission probability 
$\Gamma_j{\,}$.\par
\smallskip
\indent
Before we investigate further the probability
distribution $P[\{a_{j}\}]$ (equivalently $P[\{c_{j}\}]$), we
define, for fixed 
$j{\,}$, the measure $d^{2}a_j$ on the complex $a_{j}$-space 
to be the natural measure on $\Bbb R^2{\,}$:
$$d^{2}a_{j}{\;}{\,} 
={\;}{\,}d\bigl[{\rm Re}(a_{j})\bigr]{\;}d\bigl[{\rm Im}(a_{j})\bigr]
{\quad}.\eqno(A8)$$
\noindent
Since the probability density only depends on ${\mid}a_j{\mid}{\,}$, 
it is further convenient to rewrite the measure in terms of polar
coordinates, through $a_{j}=R_{j}{\,}e^{i\theta_j}{\,}$, 
for each $j{\,}$.  We then choose the normalisation factor $\hat N$ 
in Eqs.(4.1,2,6,A5) such that 
$${1\over\pi}{\,}\int{\;}\prod_{j}{\;}d^{2}a_{j}{\;}
P\Bigl[\{a_{j}\}\Bigr]{\;}{\,}={\;}{\,}1{\quad}.\eqno(A9)$$
\noindent
In the language of Eq.(4.6), this involves a Gaussian integral over the
infinitely many variables $a_j$ or $c_j{\,}$, which will require
regularisation, for example by the $\zeta$-function technique [31].
This procedure will be simpler when one works with a 
locally-supersymmetric theory, such as the theory of $N=1$ 
supergravity with massless-scalar/spin-$1\over 2$ supermatter [2].
In that case, there should not be any divergences in the
calculation of $\hat N$, because of cancellations between the bosonic 
and fermionic degrees of freedom, and such cancellations should make 
the detailed expression for $\hat N$ much simpler.\par
\smallskip
\indent
In Eq.(A5), the Laguerre polynomial $L_{k}(x)$ has $k$ real, distinct
roots; further, for positive argument $X_{j}>0$ and excited states
$(k_{j}>0)$, the function ${\,}e^{-X_{j}}{\;}L_{k_j}(2X_j){\,}$ takes 
negative values for certain ranges of $X_j{\,}$.  This function cannot 
therefore be interpreted as a probability density.  Regarding the
question of whether the probability distribution $P[\{a_j\}]$ might 
arise from the diagonal components of some kind of 'density matrix', let us 
consider an Hermitian position-space density matrix of the form
$$\rho(x,y){\;}{\,} 
={\;}{\,}\sum_{i}{\;}w_{i}{\;}\psi_{i}(x){\;}\psi^{*}_{i}(y)
{\quad},\eqno(A10)$$
\noindent
where the $\{\psi_{i}(x)\}$ form a complete set and $w_i$ is the
probability to be in the state $\psi_{i}(x)$.  If the system is in the
single state defined by $\psi_{i}(x)$, then the probability density
for observing $x$ is ${\mid}\psi_{i}(x){\mid}^2$.  The full
$\rho(x,y)$ does not have a probabilistic interpretation, but the 
diagonal contribution
$$P(x){\;}{\,} 
={\;}{\,}{\,}\rho(x,x){\;}{\,} 
={\;}{\,}\sum_{i}{\;}w_{i}{\;}{\,}{\mid}\psi_{i}(x){\mid}^{2}\eqno(A11)$$
\noindent
is the probability density for observing the coordinate $x{\,}$. For a
pure state, one has
$$\rho(x,y)\;=\;\Psi(x)\Psi^{*}(y)\quad.\eqno(A12)$$
\noindent 
Therefore, $P(x)$ for a pure state is the squared norm of a complex
quantum amplitude.\par
\smallskip
\indent
In our case, if the field is in the $k$-th state, then according to Eq.(A11),
${\hat N}{\,}e^{-X_j}{\;}L_{k_j}(2X_{j})$ would be the 'probability density' 
for the 'harmonic oscillator coordinate' $X_j{\,}$. Although this
function takes on negative values, the Bose-Einstein distribution
Eq.(A6) effectively smoothes out the oscillations in $\hat N
e^{-X_{j}}L_{k_{j}}(2X_{j})$ to give a positive probability density
when we sum over all possible states $\{k_{j}\}$.  Although this
provides an interesting way of breaking down the probability
distribution (A5), it does not provide us with a sensible density
matrix, derived {\it post hoc} from the quantum amplitude
$\exp\Bigl(iS^{(2)}_{\rm class}[\{a_{k\ell m}\};T]\Bigr)$, where
$S^{(2)}_{\rm class}$ is given by Eq.(A10) of [3].\par
\smallskip
\indent
The density $\hat N e^{-X_{j}}L_{k_{j}}(2X_{j})$, however,  is 
closely related to the Wigner quasi-probability distribution for the 
harmonic oscillator [32].  Wigner functions are quantum-mechanical
analogues of classical phase-space distributions, which are formally
equivalent to more familiar Hilbert-space or density-matrix
formulations of quantum mechanics. However, as the uncertainty
principle forbids the existence of a probability distribution for
simultaneously well-defined conjugate variables, such as coordinates
and momenta, Wigner functions do not have all the properties of a
classical phase-space probability density.  In particular, Wigner
functions must become negative in some domain of phase space. \par
\smallskip
\indent
For a system in a mixed state described by a density matrix
$\rho(x,y)$, the one-dimensional Wigner function is defined as [32]
$$W(p,q)\;=\; {{1}\over {2\pi \hbar}}\int^{\infty}_{-\infty}
dz\;e^{{ipz}\over{\hbar}}\;\rho\Bigl(q-{{z}\over{2}}, q
+{{z}\over{2}}\Bigr).\eqno(A13)$$
From Eq.(A10),
$$W(p,q)\;\;=\;\;\sum_{i}\,w_{i}\; W_{i}(p,q),\eqno(A14)$$ 
where
$$W_{i}(p,q)\;\;=\;\;{{1}\over{2\pi\hbar}}\int^{\infty}_{-\infty}dz\;e^{{ipz}
\over{\hbar}}\;\psi_{i}\Bigl(q-{{z}\over{2}}\Bigr)\;\psi^{*}_{i}\Bigl(q+{{z}
\over{2}}\Bigr).\eqno(A15)$$
For a pure state,
$$W(p,q)\;\;={{1}\over{2\pi\hbar}}\int^{\infty}_{-\infty}dz\;e^{{ipz}
\over{\hbar}}\;\Psi\Bigl(q-{{z}\over{2}}\Bigr)\;\Psi^{*}\Bigl(q+{{z}\over
{2}}\Bigr).\eqno(A16)$$
The $\hbar$-dependent function $W(p,q)$ is manifestly real and has the
properties
$$\eqalignno{\int^{\infty}_{-\infty}dp\;\;W(p,q)\;\;&=\;\;{\mid} \Psi(q)
{\mid}^{2},&(A17)\cr  
\int^{\infty}_{-\infty}dq\;\;W(p,q)\;\;&=\;\;{\mid}\tilde\Psi(p){\mid}^{2},
&(A18)\cr
\int^{\infty}_{-\infty} dp\;dq\;\;W(p,q)\;\;&=\;\;1,&(A19)\cr}$$
where $\tilde\Psi(p)$ is the Fourier transform of $\Psi(q)$.  For the
$k$-th state of the harmonic oscillator with the wave function
$$\psi_{k}(x)\;\;=\;\;{{1}\over{\sqrt
{2^{k}k!}}}\;\Bigl({{\omega}\over{\pi\hbar}} \Bigr)^{1\over
4}\exp\bigl(-\omega
x^{2}/2\hbar\bigr)\;H_{k}\Bigl(x\sqrt{\omega/\hbar}\Bigr),\eqno(A20)$$ 
then
$$W_{k}(p,q)\;\;=\;\;2\; (-1)^{k}\;\exp
\Bigl(-2H(p,q)/\hbar\omega\Bigr)\;L_{k}\biggl({{4H(p,q)}
\over{\hbar\omega}}\biggr),\eqno(A21)$$ 
where $H(p,q)=(1/2)(p^{2} +\omega^{2}q^{2})$ is the Hamiltonian and
$\omega$ is the oscillator frequency.  Further, using Eq.(A1),
$W_{k}(p,q)$ is normalised according to
$$\sum^{\infty}_{k=0}\;W_{k}(p,q)\;\;=\;\;1.\eqno(A22)$$
As $H$ increases, $W_{k}(p,q)$ oscillates with a decreasing amplitude
and with increasing broad peaks.  Introducing the complex variables
$\{a_{j}\}$ which represent a phase-space point $(p_{j},q_{j})$, with a
mode index $j$ for the infinite-dimensional phase space, where
$$a_{j}\;\;=\;\;{{1}\over{(2\hbar)^{1/2}}}\;\Bigl(\omega^{1/2} q_{j}
+i\omega^{-1/2} p_{j}\Bigr),\eqno(A23)$$
one can express $W_{k}(p,q)$ as
$$W_{k}(a_{j})\;\;=\;\;2\,(-1)^{k}\;e^{-2{\mid}a_{j}{\mid}^{2}}
L_{k}\Bigl(4{\mid}a_{j}{\mid}^{2}\Bigr).\eqno(A24)$$ 
Using $L_{k}(0) =1$, one has
$$W_{k}(0)\;\;=\;\;2\;(-1)^{k},\eqno(A25)$$
which is the maximum value of $W_{k}(a_{j})$.  Thus, with the
replacement $\,2{\mid}a_{j}{\mid}^{2}\rightarrow X_{j}\,$, the final term
in Eq.(A5) can be written as
$$\hat N\;e^{-X_{j}}L_{k_{j}}(2X_{j})\;\;=\;\;{1\over 4}\;\hat
N\;W_{k_{j}} (X_{j})\;W_{k_{j}}(0)\;\;=\;\;\hat
N\;{{{W_{k_{j}}(X_{j})}\over{W_{k_{j}}(0)}}}.\eqno(A26)$$
\smallskip
\indent
We can interpret the product -- or ratio -- of two Wigner functions in
the following way.  Consider two statistically independent subsystems
$A$ and $B$.  As in statistical physics, the Wigner function $W$
describing the system $A\cup B$ is given by the product of the Wigner
functions for the two subsystems
$$W(p_{A},q_{A},p_{B},q_{B})\;\;=\;\;W_{A}(p_{A},q_{A})\;W_{B}
(p_{B},q_{B}).\eqno(A27)$$ 
Thus, the state of the subsystem $B$ does not affect the
'probabilities' of various states of the other subsystem $A$.  Such a
factorisation would not be possible if the subsystems were
statistically dependent, in which case, the Wigner function of each
subsystem is defined by integrating over the other system's variables:
$$W_{A}(p_{A},q_{A})\;\;=\;\;\int\;dq_{B}\int\;dp_{B}\;\;W(p_{A},
q_{A},p_{B},q_{B}).\eqno(A28)$$
In our case, one might interpret the system $B$, say, as
corresponding to the weak initial state of the perturbations at early
times, and $A$ as corresponding to the final state of the fluctuations
at late times.  Thus, referring to Eq.(A11), the
${\mid}\psi_{i}(x){\mid}^2$ term is replaced by
$W_{k_{j}}(X_{j})/W_{k_{j}} (0)$, which gives a Wigner function for
the system $A$ at late times.\par    
\smallskip
\noindent
{\bf Acknowledgements}
\smallskip
\indent
We are grateful for numerous helpful comments from referees, which
helped clarify several matters.
\medskip
\parindent = 1 pt
{\bf References}
\medskip
\indent
[1] A.N.St.J.Farley and P.D.D'Eath, 
Phys Lett. B, {\bf 601}, 184 (2004); 
Phys Lett. B {\bf 613}, 181 (2005).\par          
\indent
[2] A.N.St.J.Farley and P.D.D'Eath, 
`Quantum Amplitudes in Black-Hole Evaporation: I. Complex Approach', 
submitted for publication (2005); 
R.P.Feynman and A.R.Hibbs, 
{\it Quantum Mechanics and Path Integrals}, 
(McGraw-Hill, New York) (1965).\par
\indent
[3] A.N.St.J.Farley and P.D.D'Eath, 
`Quantum Amplitudes in Black-Hole Evaporation: II Spin-0 Amplitude', 
submitted for publication (2005).\par
\indent      
[4] W.McLean, 
{\it Strongly Elliptic Systems and Boundary Integral Equations}, 
(Cambridge University Press, Cambridge) (2000);
O.Reula, 'A configuration space for quantum gravity and solutions to
the Euclidean Einstein equations in a slab region', 
Max-Planck-Institut f\"ur Astrophysik, {\bf MPA} 275 (1987).\par
\indent
[5] A.Das, M.Fischler and M. Ro\v cek, 
Phys. Lett. B {\bf 69}, 186 (1977).\par
\indent
[6] P.D.D'Eath, {\it Supersymmetric Quantum Cosmology}, 
(Cambridge University Press, Cambridge) (1996).\par
\indent
[7] P.D.D'Eath, 
'Loop amplitudes in supergravity by canonical quantization', in 
{\it Fundamental Problems in Classical, Quantum and String Gravity}, 
ed. N.S\'anchez (Observatoire de Paris) 166 (1999), hep-th/9807028.\par
\indent
[8] P.D.D'Eath, 
'What local supersymmetry can do for quantum cosmology', 
in {\it The Future of Theoretical Physics and Cosmology}, 
eds. G.W.Gibbons, E.P.S.Shellard and S.J.Rankin 
(Cambridge University Press, Cambridge) 693 (2003).\par
\indent
[9] L.Parker, Phys.Rev.D {\bf 12}, 1519 (1975).\par
\indent
[10] S.W.Hawking, Commun. Math. Phys. {\bf 43}, 199 (1975);
Phys. Rev. D {\bf 14}, 2460 (1976).\par
\indent
[11] N.D.Birrell and P.C.W.Davies, 
{\it Quantum fields in curved space}, 
(Cambridge University Press, Cambridge) (1982).\par
\indent 
[12] V.P.Frolov and I.D.Novikov, {\it Black Hole Physics}, 
(Kluwer Academic, Dordrecht) (1998).\par
\indent
[13] P.C.Vaidya, Proc. Indian Acad. Sci. {\bf A33}, 264 (1951); 
R.W.Lindquist, R.A.Schwartz and C.W.Misner, 
Phys. Rev. {\bf 137}, 1364 (1965); 
A.N.St.J.Farley and P.D.D'Eath, 
`Vaidya Space-Time and Black-Hole Evaporation', 
submitted for publication (2005).\par 
\indent
[14] S.W.Hawking and G.F.R.Ellis, 
{\it The large scale structure of space-time}, 
(Cambridge University Press, Cambridge) (1973).\par
\indent
[15] P.Kraus and F.Wilczek, Nucl. Phys. B {\bf 433}, 403 (1995).\par
\indent
[16] M.K.Parikh and F.Wilczek, Phys. Rev. Lett.{\bf 85}, 5042
(2000).\par
\indent
[17] J.D.Jackson, {\it Classical Electrodynamics}, 
(Wiley, New York) (1975).\par
\indent 
[18] C.W.Misner, K.S.Thorne and J.A.Wheeler, {\it Gravitation},
(Freeman, San Francisco) (1973).\par
\indent
[19] R.Arnowitt, S.Deser and C.W.Misner, 
'Dynamics of General Relativity', 
in {\it Gravitation: An Introduction to Current Research}, 
ed. L.Witten (Wiley, New York) (1962).\par
\indent
[20] J.A.H.Futterman, F.A.Handler and R.A.Matzner,   
{\it Scattering from Black Holes} 
(Cambridge University Press, Cambridge) (1988).\par
\indent 
[21] D.N.Page, Phys. Rev. D {\bf 13}, 198 (1976).\par
\indent
[22] S.W.Hawking, Phys. Rev. D {\bf 13}, 191 (1976).\par
\indent
[23] R.Casadio, B.Harms and Y. Leblanc, Phys. Rev. D {\bf 58}, 044014
(1998).\par 
\indent
[24] E.Keski-Vakkuri and P.Kraus, Nucl. Phys. B {\bf 491}, 249
(1997).\par 
\indent
[25] T.Damour and R.Ruffini, Phys. Rev. D {\bf 14}, 332 (1976).\par
\indent
[26] K.Freese, C.T.Hill and M.Mueller, Nucl. Phys. B {\bf 255}, 693
(1985).\par 
\indent
[27] A.N.St.J.Farley and P.D.D'Eath, 
Class. Quantum Grav. {\bf 22}, 2765 (2005).\par
\indent
[28] A.N.St.J.Farley and P.D.D'Eath, 
Class. Quantum Grav. {\bf 22}, 3001 (2005).\par
\indent
[29] I.S.Gradshteyn and I.M.Ryzhik, 
{\it Tables of Integrals, Series and Products} 
(Academic Press, New York) (1965).\par
\indent
[30] L.Parker, Phys. Rev. {\bf 183}, 1057 (1969).\par
\indent
[31] S.W.Hawking, Commun. Math. Phys. {\bf 56}, 133 (1977).\par
\indent
[32] M.H.Hillery, R.F.O'Connell, M.O.Scully and E.P.Wigner,
Phys. Rep. {\bf 106}, 121 (1984).\par
\indent

\end